\documentclass[seceq,preprint]{ptptex}

\title{
Lovelock Gravity with Spontaneous Dimensional Breaking
}

\author{
Kouzou \textsc{Nishida}%
\footnote{E-mail: EZF01671@nifty.com} 
}

\inst{
Department of Physics, Kyoto Sangyo University, Kyoto 603-8555, Japan 
}

\abst{
We consider $D$-dimensional Lovelock gravity 
with only one term of higher-order Lovelock Lagrangian densities, 
and show that a product of Minkowski space-time and $n$-spheres 
is its vacuum solution.
The most interesting feature of our model is that 
the spontaneous compactification of the extra dimensions 
results in reproduction of the Einstein gravity with no cosmological constant.
}

\begin{document}

\maketitle

\section{Introduction}
Lovelock gravity\cite{rf:1} is a natural extension of Einstein gravity to higher dimensions, 
and it is of great interest in theoretical physics as it describes a wide class of models.
The equations of motion of Lovelock gravity do not contain more than second 
order derivatives with respect to metric, as in the case of general relativity.

Recently, Maeda and Dadhich\cite{rf:2} presented vacuum solutions of the form 
${\cal M}^m \times {\cal H}^n $ in Gauss Bonnet gravity theory. 
Cai, Cao, and Ohta\cite{rf:3} generalized the solutions in $D$-dimensional Lovelock gravity. 
Here, ${\cal M}^m$ is the $m$-dimensional manifold 
and ${\cal H}^n$ is a negative constant curvature space.

In this paper, we consider $D$-dimensional Lovelock gravity 
with only one term of higher-order Lovelock Lagrangian densities,
and show that a product of Minkowski space-time 
and $n$-spheres (positive constant curvature space) is its vacuum solution.
This vacuum solution indicates the possibility that 
compactification of the extra dimensions takes place spontaneously.
We show that the usual Einstein gravity with no cosmological constant 
is reproduced as a result of the spontaneous compactification.

To illustrate our idea, let us consider the simplest example. 
We take the following Lagrangian
\begin{equation}
\label{101}
S=\frac{1}{2\kappa^2}\int d^{6}x\sqrt{-g}{\cal L}^{(6)}_{GB},
\end{equation}
where ${\cal L}^{(6)}_{GB}$ is the six-dimensional Gauss-Bonnet term 
(second order Lovelock Lagrangian density)
${\cal L}^{(6)}_{GB}\equiv R_{\mu\nu\rho\sigma}R^{\mu\nu\rho\sigma}-4R_{\mu\nu}R^{\mu\nu}+R^{2}$.
Let us compactify the six-dimensional space-time to a product of a
four-dimensional manifold and a two-sphere, that is, ${\cal M}^D \rightarrow {\cal M}^4 \times S^2$.
By substituting $R_{abcd}=r^2(\eta_{ac}\eta_{bd}-\eta_{ad}\eta_{bc})$
in the two-sphere part,
eqn. (\ref{101}) can be rewritten as
\begin{equation}
\label{102}
S=\frac{V^{(2)}}{2\kappa^2}\int d^{4}x\sqrt{-g}\frac{4}{r^2}R^{(4)},
\end{equation}
where $V^{(2)}$ represents the volume of $S^2$ and 
we have used $\int d^{4}x\sqrt{-g}{\cal L}^{(4)}_{GB}=\int d^{4}x\sqrt{-g}{\cal L}^{(2)}_{GB}=0$.
Note that we obtained the Einstein-Hilbert action
and (\ref{102}) does not include the cosmological constant.

This example could not reproduce the Standard Model, 
because it could only yield the gauge symmetry of $SU(2)$, which comes from $S^2$, 
when compactified down to four dimensions.
In the next section, we generalize the above treatment to higher dimensions.

\section{The $D$-dimensional model}
We consider the following Lagrangian
\begin{equation}
\label{201}
S=\frac{1}{2\kappa^{2}_{(D)}}\int d^{D}x\sqrt{-g}{\cal L}^{(D)}_{(p)}+S_{\rm matter},
\end{equation}
where $p \leq [(D-1)/2]$
 ($[N]$ denotes the integral part of the number $N$),
$\kappa_{(D)} \equiv \sqrt{8\pi G_{(D)}}$ where $G_{(D)}$ is
the $D$-dimensional gravitational constant
and ${\cal L}^{(D)}_{(p)}$ is the $D$-dimensional $p$-th order Lovelock Lagrangian density
\begin{equation}
\label{202}
{\cal L}^{(D)}_{(p)}
\equiv
\frac{1}{2^{p}}
\delta^{\mu_1\nu_1 \cdots \mu_p\nu_p}_{\lambda_1\sigma_1 \cdots\lambda_p\sigma_p}
R^{\lambda_1\sigma_1}\,_{\mu_1\nu_1}\cdots R^{\lambda_p\sigma_p}\,_{\mu_p\nu_p}. 
\end{equation}
The $\delta$ symbol denotes a totally anti-symmetric product of
Kronecker deltas, normalized to take values 0 and $\pm 1$.

Varying the action (\ref{201}), we obtain the equations of motion as
\begin{equation}
\label{203}
G^{(D)}_{(p)\,\mu\nu}=\kappa^{2}_{(D)}T_{\mu\nu},
\end{equation}
where $T_{\mu\nu}$ is the energy-momentum tensor of the matter field 
derived from $S_{\rm matter}$ and
\begin{eqnarray}
\label{204}
G^{(D)\mu}_{(p)\:\:\:\:\nu}
& \equiv & 
\delta^{\mu_1\nu_1 \cdots \mu_p\nu_p}_{\nu\sigma_1 \cdots\lambda_p\sigma_p}
R^{\mu\sigma_1}\,_{\mu_1\nu_1}\cdots R^{\lambda_p\sigma_p}\,_{\mu_p\nu_p} 
-\frac{1}{2}\delta^{\mu}_{\nu}{\cal L}^{(D)}_{(p)}
\nonumber \\
& = &
-\frac{1}{2^{p+1}}
\delta^{\mu\mu_1\nu_1 \cdots \mu_p\nu_p}_{\nu\lambda_1\sigma_1
\cdots\lambda_p\sigma_p}
R^{\lambda_1\sigma_1}\,_{\mu_1\nu_1}\cdots R^{\lambda_p\sigma_p}\,_{\mu_p\nu_p}. 
\end{eqnarray}

Let us compactify the $D$-dimensional space-time to
a product of an $m$-dimensional manifold and an $n$-sphere, 
that is, ${\cal M}^{D} \rightarrow {\cal M}^{m} \times S^n$.
For $S^n$, the Riemann curvature can be calculated as
\begin{equation}
\label{205}
R_{abce}=r_{(n)}^2(\eta_{ac}\eta_{bd}-\eta_{ad}\eta_{bc}),
\,\,\,\,\,\,
R^{ab}\,_{cd}=\frac{1}{r_{(n)}^2}\delta^{ab}_{cd},
\end{equation}
where $r_{(n)}$ is the constant radius of the $n$-sphere.
We indicate the vectors in the $m$-dimensional manifold by subscripted
Roman letters $i$, $j$, $k$, etc. and the vectors in the $n$-sphere by subscripted
Roman letters $a$, $b$, $c$, etc. 
Greek letters such as $\mu$ and $\nu$ are used to label $D$-dimensional space-time vectors.
Substituting (\ref{205}) into (\ref{204}),
we can decompose (\ref{204}) into $m$-dimensional and $n$-dimensional parts\cite{rf:3}:
\begin{eqnarray}
\label{206}
G^{(D)i}_{(p)\:\:\:\:j}
& = &
-\frac{1}{2^{p+1}}
\delta^{i\mu_1\nu_1 \cdots \mu_p\nu_p}_{j\lambda_1\sigma_1
\cdots\lambda_p\sigma_p}
R^{\lambda_1\sigma_1}\,_{\mu_1\nu_1}\cdots R^{\lambda_p\sigma_p}\,_{\mu_p\nu_p} 
\nonumber \\
& = &
-\sum^{p}_{t=0}
\left(
\begin{array}{l}
p \\
t
\end{array}
\right)
\frac{1}{2^{p+1}}
\delta^{i k_1 l_1 \cdots k_t l_t c_1 d_1 \cdots c_{p-t} d_{p-t}}
      _{j m_1 n_1 \cdots m_t n_t e_1 f_1 \cdots e_{p-t} f_{p-t}}
R^{m_1 n_1}\,_{k_1 l_1}\cdots R^{m_t n_t}\,_{k_t l_t} 
\nonumber \\
& &
\times
R^{e_1 f_1}\,_{c_1 d_1}\cdots R^{e_{k-t} f_{k-t}}\,_{c_{k-t} d_{k-t}} 
\nonumber \\
& = &
\sum^{[(m-1)/2]}_{t=0}
\left(
\begin{array}{l}
p \\
t
\end{array}
\right)
\frac{n!}{(n-2(p-t))!}\left(\frac{1}{r_{(n)}^2} \right)^{p-t} \times
G^{(m)i}_{(t)\:\:\:\:j},
\end{eqnarray}
\begin{equation}
\label{207}
G^{(m)i}_{(t)\:\:\:\:j}
\equiv
-\frac{1}{2^{t+1}}
\delta^{ik_1 l_1 \cdots k_t l_t}
      _{jm_1 n_1 \cdots m_t n_t}
R^{m_1 n_1}\,_{k_1 l_1}\cdots R^{m_t n_t}\,_{k_t l_t}, 
\end{equation}
\begin{eqnarray}
\label{208}
G^{(D)a}_{(p)\:\:\:\:b}
& = &
-\frac{1}{2^{p+1}}
\delta^{a\mu_1\nu_1 \cdots \mu_p\nu_p}_{b\lambda_1\sigma_1 \cdots\lambda_p\sigma_p}
R^{\lambda_1\sigma_1}\,_{\mu_1\nu_1}\cdots R^{\lambda_p\sigma_p}\,_{\mu_p\nu_p} 
\nonumber \\
& = &
-\sum^{p}_{t=0}
\left(
\begin{array}{l}
p \\
t
\end{array}
\right)
\frac{1}{2^{p+1}}
\delta^{a k_1 l_1 \cdots k_t l_t c_1 d_1 \cdots c_{k-t} d_{k-t}}
      _{b m_1 n_1 \cdots m_t n_t e_1 f_1 \cdots e_{k-t} f_{k-t}}
R^{m_1 n_1}\,_{k_1 l_1}\cdots R^{m_t n_t}\,_{k_t l_t} 
\nonumber \\
& &
\times
R^{e_1 f_1}\,_{c_1 d_1}\cdots R^{e_{k-t} f_{k-t}}\,_{c_{k-t} d_{k-t}} 
\nonumber \\
& = &
-\frac{1}{2}\delta^{a}_{b}
\sum^{[m/2]}_{t=0}
\left(
\begin{array}{l}
p \\
t
\end{array}
\right)
\frac{(n-1)!}{(n-1-2(p-t))!}\left(\frac{1}{r_{(n)}^2} \right)^{p-t}
\times {\cal L}^{(m)}_{(t)},
\end{eqnarray}
\begin{equation}
\label{209}
{\cal L}^{(m)}_{(t)}
\equiv
\frac{1}{2^{t}}
\delta^{k_1 l_1 \cdots k_t l_t}
      _{m_1 n_1 \cdots m_t n_t}
R^{m_1 n_1}\,_{k_1 l_1}\cdots R^{m_t n_t}\,_{k_t l_t}, 
\end{equation}
where $p-t \geq 1$, and we have used the identities for the Lovelock tensors\cite{rf:2}:
\begin{equation}
\label{210}
G^{(m)i}_{(t)\:\:\:\:j}\equiv 0
\,\,\,
{\rm for}
\,\,\,
m \leq 2t,
\,\,\,
{\cal L}^{(m)}_{(t)}\equiv 0
\,\,\,
{\rm for}
\,\,\,
m \leq 2t-1,
\end{equation}
and the identity
\begin{equation}
\label{211}
\delta^{\mu_1 \cdots \mu_{p-1}\mu_p}
      _{\nu_1 \cdots \nu_{p-1}\nu_p}
\delta^{\mu_{p-1}\mu_p}
      _{\nu_{p-1}\nu_p}
=2[r-(p-1)][r-(p-2)]
\delta^{\mu_1 \cdots \mu_{p-2}}
      _{\nu_1 \cdots \nu_{p-2}}
\,\,\,
(p \geq 2),
\end{equation}
where $r$ denotes the range of the index
($r=m$ for ${\cal M}^m$ and $r=n$ for $S^n$)
and $\delta^{\mu_1 \cdots \mu_{p-2}} _{\nu_1 \cdots \nu_{p-2}} \equiv 1$ for $p=2$.
The other components (such as $G^{(m)a}_{(t)\:\:\:\:i}$) automatically vanish.

The Lagrangian (\ref{202}) can be decomposed as 
\begin{equation}
\label{212}
{\cal L}^{(D)}_{(p)\:\:\:\:}
=
\sum^{[m/2]}_{t=0}
\left(
\begin{array}{l}
p \\
t
\end{array}
\right)
\frac{n!}{(n-2(p-t))!}\left(\frac{1}{r_{(n)}^2} \right)^{p-t}
\times {\cal L}^{(m)}_{(t)},
\end{equation}
where $p-t \geq 1$.

Let us consider the most important case ${\cal M}^4 \times  S^{n}$.
By substituting $m=4$ in (\ref{206}), (\ref{208}), and (\ref{212}), we obtain
\begin{eqnarray}
\label{213}
G^{(D)i}_{(p)\:\:\:\:j}
& = &
n(n-1)\cdots (n-2p+1)\left( \frac{1}{r_{(n)}^2} \right)^{p}
G^{(4)i}_{(0)\:\:\:\:j}
\nonumber \\
& &
+
n(n-1)\cdots (n-2p+3)\left( \frac{1}{r_{(n)}^2} \right)^{p-1}
G^{(4)i}_{(1)\:\:\:\:j},
\end{eqnarray}
\begin{eqnarray}
\label{214}
G^{(D)a}_{(p)\:\:\:\:b}
& = &
-\frac{1}{2}\delta^{a}_{b}
\left[
(n-1)\cdots (n-2p)\left( \frac{1}{r_{(n)}^2} \right)^{p}
{\cal L}^{(4)}_{(0)} 
\right.
\nonumber \\
& &
+
p(n-1)\cdots (n-2p+2)\left( \frac{1}{r_{(n)}^2} \right)^{p-1}
{\cal L}^{(4)}_{(1)} 
\nonumber \\
& &
+
\left.
\frac{p(p-1)}{2}
(n-1)\cdots (n-2p+4)\left( \frac{1}{r_{(n)}^2} \right)^{p-2}
{\cal L}^{(4)}_{(2)} 
\right],
\end{eqnarray}
\begin{eqnarray}
\label{215}
{\cal L}^{(D)}_{(p)}
& = &
n(n-1)\cdots (n-2p+1)\left( \frac{1}{r_{(n)}^2} \right)^{p}
{\cal L}^{(4)}_{(0)} 
\nonumber \\
& &
+
pn(n-1)\cdots (n-2p+3)\left( \frac{1}{r_{(n)}^2} \right)^{p-1}
{\cal L}^{(4)}_{(1)} 
\nonumber \\
& &
+
\frac{p(p-1)}{2}
n(n-1)\cdots (n-2p+5)\left( \frac{1}{r_{(n)}^2} \right)^{p-2}
{\cal L}^{(4)}_{(2)}. 
\end{eqnarray}
Note that the usual Einstein gravity with no cosmological constant can be reproduced
when
\begin{equation}
\label{216}
n+1 \leq 2p <  n+3.
\end{equation}
For example, when ${\cal M}^{13} \rightarrow {\cal M}^4 \times S^9$
($n=9$) and $p=5$,
by using (\ref{213}) and (\ref{214}),  
the equations of motion (\ref{203}) can be decomposed as  
\begin{equation}
\label{217}
R^{(4)}_{ij}-\frac{1}{2}g_{ij}R^{(4)} =\kappa^{2}_{(4)}T_{ij}
\end{equation}
for the four-dimensional parts, and
\begin{equation}
\label{218}
-\delta^a_b\left(R^{(4)}+r_{(9)}^2 {\cal L}_{(2)}^{(4)}\right)
=18\kappa_{(4)}^{2}T^{a}\,_{b}
\end{equation}
for the nine-dimensional parts.
By using (\ref{215}), (\ref{201}) can be rewritten as
\begin{equation}
\label{219}
S=\frac{V^{(n)}}{2\kappa^{2}_{(4)}}\int d^{4}x\sqrt{-g}R^{(4)}+S_{\rm matter},
\end{equation}
where $V^{(n)}$ is the volume of $S^n$ and 
\begin{equation}
\label{220}
\kappa^{2}_{(4)} \equiv \frac{r_{(9)}^8}{5 \times 9!}\kappa^{2}_{(13)}. 
\end{equation}
Incidentally, the space-time ${\cal M}^4 \times S^9$ has a gauge symmetry 
of $SO(10)$ which comes from $S^9$.

In the case of vacuum ($T_{\mu\nu}=0$), Minkowski space-time is a trivial
solution of (\ref{217}) and (\ref{218}), namely, 
the product of a four-dimensional Minkowski and 
an $n$-sphere is a solution of (\ref{203}).
This vacuum solution indicates the possibility that 
compactification of the extra dimensions takes place spontaneously.

Next, let us consider the compactification 
${\cal M}^{13} \rightarrow {\cal M}^4 \times S^{2} \times S^{'2} \times S^5$. We set $p=5$. 
Because $SO(3)=SU(2)$ and $SO(6)=SU(4)$,
the space-time ${\cal M}^4 \times S^{2} \times S^{'2} \times S^5$ 
has the gauge symmetry of $SU(4)\otimes SU(2)\otimes SU(2)$ as the internal symmetry.
First, we compactify ${\cal M}^{13}$ to ${\cal M}^8 \times S^5$.
By using (\ref{206}),
the eight-dimensional parts of $G^{(13)}_{(5)\mu\nu}$ is 
decomposed as 
\begin{equation}
\label{221}
G^{(13)}_{(5)ij}=
\frac{10 \times 5!}{r_{(5)}^4}
G^{(8)}_{(3)ij}.
\end{equation}
Next, we compactify ${\cal M}^8$ to ${\cal M}^6 \times S^{'2}$. 
Then the six-dimensional parts of $ G^{(8)}_{(3)ij}$ are 
decomposed as 
\begin{equation}
\label{222}
G^{(8)}_{(3)ij}=
\frac{6}{r_{(2)}^{'2}}
G^{(6)}_{(2)ij}.
\end{equation}
Lastly, we compactify ${\cal M}^6$ to ${\cal M}^4 \times S^{2}$.
The four-dimensional parts of $G^{(6)}_{(2)ij}$ are 
decomposed as 
\begin{equation}
\label{223}
G^{(6)}_{(2)ij}=
\frac{4}{r_{(2)}^{2}}
G^{(4)}_{(1)ij}.
\end{equation}
Combining (\ref{221}), (\ref{222}), and (\ref{223}), 
we obtain the four-dimensional parts of $G^{(13)}_{(5)\mu\nu}$ as 
\begin{equation}
\label{224}
G^{(13)}_{(5)ij}
=
\frac{28800}{r_{(5)}^4 r_{(2)}^{'2} r_{(2)}^2}
G_{(1)ij}^{(4)}.
\end{equation}
Similarly,
by using (\ref{208}) and (\ref{212}), 
the five-dimensional parts of $G^{(13)}_{(5)\mu\nu}$ can be decomposed as  
\begin{equation}
\label{225}
G^{(13)a}_{(5)\:\:\:\:b}
=
-\frac{2880}{r_{(5)}^4 r_{(2)}^{'2} r_{(2)}^2}
\delta^a_b
\left({\cal L}_{(1)}^{(4)}+\frac{r_{(5)}^2}{2} {\cal L}_{(2)}^{(4)}\right)
\end{equation}
and the two two-dimensional parts of $G^{(13)}_{(5)\mu\nu}$can be decomposed as
\begin{equation}
\label{226}
G^{(13)}_{(2)xy}
=0,
\,\,\,\,\,\,
G^{(13)}_{(2)x'y'}
=0,
\end{equation}
where the vectors in $S^5$ are indicated by subscripted Roman letters $a$, $b$, etc.; 
the vectors in $S^{2}$ by subscripted Roman letters $x$, $y$, etc.; and 
the vectors in $S^{'2}$ by subscripted Roman letters $x'$, $y'$, etc. 
The equations of motion (\ref{203}) can therefore be rewritten as
\begin{equation}
\label{227}
R^{(4)}_{ij}-\frac{1}{2}g_{ij}R^{(4)}=\kappa_{(4)}^{2}T_{ij},
\end{equation}
\begin{equation}
\label{228}
-\delta^{a}_{b}
\left(R^{(4)}+\frac{r_{(5)}^2}{2} {\cal L}_{(2)}^{(4)}\right)
R^{(4)}=10\kappa_{(4)}^{2}T^{a}\,_{b},
\end{equation}
\begin{equation}
\label{229}
0=T_{xy},
\,\,\,\,\,\,
0=T_{x'y'},
\end{equation}
where
\begin{equation}
\label{230}
\kappa^{2}_{(4)} \equiv \frac{r_{(5)}^4 r_{(2)}^{'2} r_{(2)}^{2}}
{28800}
\kappa^{2}_{(13)}. 
\end{equation}
Thus, we have retrieved the Einstein gravity with no cosmological constant.

\section{Conclusion}

In this paper, we considered $D$-dimensional Lovelock gravity 
with only one term of the higher-order Lovelock Lagrangian densities.
We show that a product space of Minkowski space-time and $n$-spheres is its vacuum solution.
Furthermore, we showed that the Einstein gravity with no cosmological constant 
can be obtained as a result of spontaneous compactification of the extra dimensions. 
It is remarkable that without introducing negative constant curvature spaces 
and adjusted coefficients, we were able to obtain the compactified vacuum solutions 
and the zero cosmological constant.


%

\end{document}